\documentclass[11pt,twoside,a4paper,cmspaper,final,collab]{cms-tdr}
\def\svnVersion{cf69b4c-D}\def\svnDate{2026/11/06}\def\cmsCernNoTag{CERN-EP-2026-251}\def\cmsCernDate{\today}\def\cmsMessage{Submitted to Physical Review D}
\begin{document}\cmsNoteHeader{B2G-24-005}

\newcommand{\DeepJet}{\ensuremath{\textsc{DeepJet}}\xspace}
\newcommand{\WW}{\ensuremath{\PW\PW}\xspace}
\newcommand{\WZ}{\ensuremath{\PW\PZ}\xspace}
\newcommand{\ZZ}{\ensuremath{\PZ\PZ}\xspace}
\newcommand{\ttH}{\ensuremath{\ttbar\PH}\xspace}
\newcommand{\ttW}{\ensuremath{\ttbar\PW}\xspace}
\newcommand{\ttZ}{\ensuremath{\ttbar\PZ}\xspace}
\newcommand{\ST}{\ensuremath{S_{\mathrm{T}}}\xspace}
\newcommand{\Zjets}{\ensuremath{\PZ\text{ + jets}}\xspace}
\newcommand{\Wjets}{\ensuremath{\PW\text{+ jets}}\xspace}
\newcommand{\HTj}{\ensuremath{\HT^{3j}}\xspace}
\newcommand{\HTb}{\ensuremath{\HT^{\PQb}}\xspace}

\ifthenelse{\boolean{cms@external}}{\providecommand{\cmsLeft}{upper\xspace}}{\providecommand{\cmsLeft}{left\xspace}}
\ifthenelse{\boolean{cms@external}}{\providecommand{\cmsRight}{lower\xspace}}{\providecommand{\cmsRight}{right\xspace}}
\newlength{\cmsTabSkip}\setlength{\cmsTabSkip}{2ex}

\cmsNoteHeader{B2G-24-005}
\title{Search for heavy top quark partners with charge 5/3 in the single-lepton final state in proton-proton collisions at \texorpdfstring{$\sqrt{s} = 13\TeV$}{sqrt(s) = 13 TeV}}

\author[cern]{The CMS Collaboration}

\date{\today}

\abstract{
A search for the pair production of a vector-like quark, $\PX_{5/3}$, a heavy fermionic partner of the top quark with an electric charge of 5/3, is presented.  The search is performed in the final state with one lepton, an electron or a muon.  Proton-proton collision data collected by the CMS experiment at the LHC from 2016 to 2018, at a center-of-mass energy of 13\TeV, corresponding to an integrated luminosity of 138\fbinv, are used.  Two decay modes are considered.  In the conventional decay mode, the $\PX_{5/3}$ is assumed to decay exclusively into a top quark and a \PW boson.  In an alternative, exotic scenario, it is assumed to decay exclusively into a top quark and a charged Higgs boson, assumed to subsequently decay exclusively into top and bottom quarks.  No excess of data events over the standard model (SM) expectation is observed.  Expected and observed upper limits on the pair production cross section of the $\PX_{5/3}$ are set at 95\% confidence level as functions of the $\PX_{5/3}$ mass.  The observed exclusion ranges of the $\PX_{5/3}$ mass extend up to 1630\GeV in the decay to SM particles, and up to 1740\GeV in the exotic decay mode for a charged Higgs boson mass of 800\GeV. These are the most stringent limits on the $\PX_{5/3}$ decaying to a top quark and a \PW boson, and the first limits on the $\PX_{5/3}$ decay into a top quark and a charged Higgs boson.}

\hypersetup{%
pdfauthor={CMS Collaboration},%
pdftitle={Search for heavy top quark partners with charge 5/3 in the single-lepton final state in proton-proton collisions at sqrt{s}=13 TeV},%
pdfsubject={CMS},%
pdfkeywords={CMS, physics, vector-like quarks}} % 

\maketitle

\section{Introduction}
\label{sec:Intro}
Vector-like quarks (VLQs) are hypothetical spin-1/2 fermions whose left- and right-handed components transform identically under the standard model (SM) electroweak (EW) gauge symmetry group.
They are predicted in various beyond-the-SM theories, including little Higgs and composite Higgs models that attempt to address the hierarchy problem~\cite{Schmaltz:2002wx,PhysRevD.75.055014,compHiggs}.
This problem arises from quadratic divergences in quantum loop corrections when the Higgs boson mass is renormalized~\cite{Matsedonskyi2013}.
By introducing VLQs, these models stabilize the quantum corrections to the Higgs boson mass.
In particular, they predict new color-triplet partners of SM quarks, including the $\PX_{5/3}$, a heavy fermionic partner of the top quark, \PQt, with an electric charge of 5/3 times the charge of the positron. 
In what follows, charge-conjugate states are implied.

\begin{figure}[hbtp]
\centering
	\includegraphics[width=0.49\textwidth]{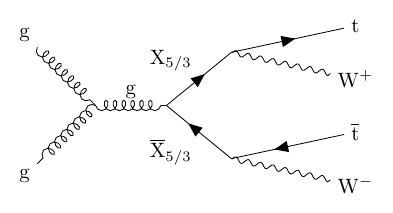}
	\includegraphics[width=0.49\textwidth]{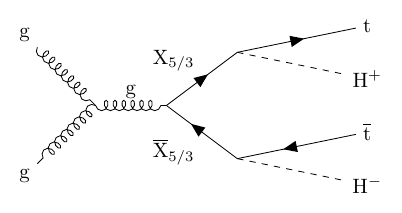}
	\caption{Representative  Feynman diagrams of $\PX_{5/3}$ pair production and their subsequent decays.}\label{fig:X53BSM}
\end{figure}

In this paper, we present a search with the CMS experiment at the CERN LHC for pair production of the $\PX_{5/3}$, which proceeds primarily via quantum chromodynamics (QCD) processes, as illustrated in Fig.~\ref{fig:X53BSM}. 
The search uses proton-proton ($\Pp\Pp$) data collected from 2016 to 2018, corresponding to an integrated luminosity of 138\fbinv. 
Searches for $\PX_{5/3}$ have traditionally assumed that it decays exclusively into a \PW boson and a top quark, as shown in Fig.~\ref{fig:X53BSM} (left)~\cite{COMPH,Cacciapaglia_2013}. This decay mode is referred to as the conventional decay mode. 
However, exotic decays of the $\PX_{5/3}$ are motivated by extensions of the SM.
These decays can proceed via $\PX_{5/3} \to \PSHp \PQt$, where $\PSHp$ denotes a charged Higgs boson whose mass is treated as a free parameter in the range 200\mbox{--}1000\GeV~\cite{Corcella:2021mdl}. 
Therefore, an alternative $\PX_{5/3}$ decay mode into $\PQt\PSHp$ is explored as shown in Fig.~\ref{fig:X53BSM} (right). 
This decay mode is referred to as the exotic decay mode. 
The charged Higgs boson is assumed to decay exclusively to a top quark and a bottom quark, \PQb.
This analysis presents the first experimental limits on the $\PX_{5/3}$ VLQ decaying into a top quark and an $\PSHp$.

Previous searches for pair-produced $\PX_{5/3}$ assuming the conventional decay mode have been conducted by the CMS and ATLAS Collaborations in the single-lepton and same-sign dilepton final states, with the data found to be compatible with the SM expectation~\cite{X5/3,Sirunyan2019-eo,Aad_2023}. 
These channels provide complementary sensitivity where the same-sign dilepton channel benefits from reduced SM backgrounds, while the single-lepton channel offers higher signal yields and acceptance. 
Previous searches have also considered both pair and single production of the $\PX_{5/3}$~\cite{Sirunyan2019-eo,X5/3}. Single production depends on the EW couplings of the $\PX_{5/3}$ and can become increasingly important at high $\PX_{5/3}$ masses~\cite{Sirunyan2019-eo}. 
The CMS Collaboration used the 2016 data to set upper limits at 95\% confidence level (\CL) on the $\PX_{5/3}$ pair production cross section for $\PX_{5/3}$ masses up to about 1300\GeV~\cite{X5/3,Sirunyan2019-eo}. 
The CMS Collaboration also performed searches in signatures with multiple top quarks that could potentially provide sensitivity to $\PX_{5/3}$ pair production through reinterpretation~\cite{CMS:2026zfh}. 
Although not optimized specifically for $\PX_{5/3}$ pair production, these searches provide a complementary probe of scenarios with final-state signatures similar to those considered in this analysis. 

Each top quark decays to a \PW boson and a bottom quark, and each \PW boson can decay either hadronically or leptonically.
In the conventional decay mode, $\PX_{5/3}$ pair production results in a signature with four \PW bosons and two bottom quarks.
In the exotic decay mode, the presence of four top quarks leads to a signature with four \PW bosons and six bottom quarks.
This analysis targets events with one electron or muon in the final state, in which one \PW boson decays leptonically and the remaining \PW bosons decay hadronically. 
This channel provides high signal acceptance and background rejection. 
The final state is characterized by one isolated charged lepton, $\Pell$, either an electron or a muon, missing transverse momentum (\ptvecmiss) from the neutrino, and multiple hadronically decaying top quarks and \PW bosons, including several jets originating from bottom quarks.

The narrow-width approximation (NWA) simplifies scattering calculations by treating unstable particles as on-shell states, under the assumption that their decay width is much smaller than their mass.
Previous CMS studies performed using the 2016 data set investigated both left- and right-handed coupling scenarios and found the resulting kinematic distributions and statistical interpretation  to be largely similar~\cite{X5/3}. There were no significant differences observed in the expected sensitivity. 
These studies indicate that the dependence on the chiral structure is negligible within the analysis precision. 
This analysis follows the same theoretical framework, in which the NWA is used to factorize the $\PX_{5/3}$ cross section into independent production and decay components, allowing the decay of the resonance to be treated independently of the production process.
Motivated by the results of the previous study, and within the signal generation setup used in this analysis, only the right-handed coupling is considered to reduce the dimensionality of the parameter space.

This paper is organized as follows.
Section~\ref{sec:CMS} provides a description of the CMS detector.
Section~\ref{sec:Samples} describes the data and simulated samples. 
Section~\ref{sec:Object} describes the reconstruction of physics objects, while Section~\ref{sec:Selection} details the event selection.
Section~\ref{sec:Systematics} presents the systematic uncertainties, followed by Section~\ref{sec:Statistics} that describes the results. 
The paper is summarized in Section~\ref{sec:Conclusion}. Tabulated results are available in the HEPData record~\cite{hepdata} for this analysis.

\section{The CMS detector}
\label{sec:CMS}
The CMS apparatus~\cite{CMS:2008xjf,CMS:2023gfb} is a multipurpose, nearly hermetic detector designed to trigger on~\cite{CMS:2020cmk,CMS:2016ngn,CMS:2024aqx} and identify electrons, photons, muons, and charged and neutral hadrons~\cite{CMS:2018rym,CMS:2020uim,CMS:2014pgm}.
Its central feature is a superconducting solenoid of 6\unit{m} internal diameter, providing a magnetic field of 3.8\unit{T}.
Within the solenoid volume are a silicon pixel and strip tracker, a lead tungstate crystal electromagnetic calorimeter (ECAL), and a brass and scintillator hadron calorimeter (HCAL), each composed of a barrel and two endcap sections.
Forward calorimeters extend the pseudorapidity ($\eta$) coverage provided by the barrel and endcap detectors.
Muons are reconstructed using gas-ionization detectors embedded in the steel flux-return yoke outside the solenoid.
More detailed descriptions of the CMS detector, together with a definition of the coordinate system used and the relevant kinematic variables, can be found in Refs.~\cite{CMS:2008xjf,CMS:2023gfb}.

Events of interest are selected using a two-tiered trigger system.
The first level (L1), composed of custom hardware processors, uses information from the calorimeters and muon detectors to select events at a rate of around 100\unit{kHz} within a fixed latency of 4\mus~\cite{CMS:2020cmk}.
The second level, known as the high-level trigger, consists of a farm of processors running a version of the full event reconstruction software optimized for fast processing, and reduces the event rate to a few\unit{kHz} before data storage~\cite{CMS:2016ngn,CMS:2024aqx}.

\section{Data and simulated samples}
\label{sec:Samples}
The analysis uses $\Pp\Pp$ collision data collected during the 2016--2018 LHC operation periods, corresponding to a total integrated luminosity of 138\fbinv.
The Monte Carlo (MC) simulation of signal and background processes is performed using several matrix element generators.

The MC signal simulation of pair production of the $\PX_{5/3}$ and its subsequent decays is performed at leading order (LO) using the \MGvATNLO~2.6.5~\cite{MADGRAPH} generator.
Signal events are generated for the right-handed coupling scenario, with a decay width below 1\% of the $\PX_{5/3}$ mass to ensure the validity of the NWA.
Two decay modes of the $\PX_{5/3}$ are considered separately:
\begin{description}
    \item[conventional decay mode:] Signal samples with $\PX_{5/3}$ masses ranging from 700--1700\GeV are simulated in 100\GeV increments.
    Each $\PX_{5/3}$ is pair produced and decays into a top quark and a \PW boson with a 100\% branching fraction.
    \item[exotic decay mode:] Signal samples with $\PX_{5/3}$ masses ranging from 600--1700\GeV are simulated in 100\GeV increments. An additional signal sample is generated at 1900\GeV to extend the high-mass coverage of the analysis, particularly for the unexplored exotic decay mode. Each $\PX_{5/3}$ is pair produced and decays into a top quark and an $\PSHp$ with a 100\% branching fraction~\cite{Corcella:2021mdl}.
    The charged Higgs boson subsequently decays into a top quark and a bottom quark with a 100\% branching fraction.
    The \PHpm mass is varied from 200--1000\GeV in 200\GeV increments, where kinematically allowed.
\end{description}

The $\PX_{5/3}$ mass spacing of 100\GeV is consistent with that used in previous searches and provides sufficient granularity to capture variations in signal acceptance across the mass spectrum.   
 The theoretical prediction for the $\PX_{5/3}$ pair production cross section depends only on the $\PX_{5/3}$ mass and is calculated at next-to-next-to-leading order (NNLO) using the \textsc{Top++} 2.0 program~\cite{Czakon:2011xx,X53xsec_paper}.

The relevant SM background processes considered include top quark-antiquark pair (\ttbar) production, \Wjets, QCD multijet, single $\PQt$, Drell--Yan, \WW, \WZ, \ZZ, \ttW, \ttZ, and \ttH events.
The \POWHEG 2.0~\cite{POWHEG1,POWHEG2,POWHEG3,POWHEG4} event generator is used to simulate \ttH, \ttbar, and single $\PQt$ production in the t channel and in association with a \PW boson at next-to-leading order (NLO) accuracy in perturbative QCD.
The \MGvATNLO generator is used to simulate Drell--Yan, \Wjets, $s$-channel single $\PQt$, \ttZ, \ttW, and QCD multijet events.
The $s$-channel single \PQt and \ttZ samples are generated at NLO.
For the \ttW process, the FxFx~\cite{FxFx} prescription is used to merge multiparton processes at NLO.
The \Wjets and Drell--Yan samples are generated at LO with MLM~\cite{MLM} merging.
The LO \PYTHIA~8.240 generator~\cite{pythia} is used to simulate vector boson pair production events. 

For all samples, \PYTHIA is used to model parton showering, hadronization, and the underlying event description.
The parton distribution functions (PDFs) are taken from the NNPDF3.1 set at NNLO~\cite{NNPDF:2017mvq}, and the underlying event tune is CP5~\cite{Tune}.
The CMS detector response is simulated using \GEANTfour~\cite{GEANT4}.
The effects of additional $\Pp\Pp$ interactions in the same or adjacent bunch crossings (pileup) are included in all simulated samples.
Simulated events are reweighted to reproduce the mean number of interactions per bunch crossing observed in data.

Background processes are grouped into several categories.
The dominant contribution arises from the \ttbar background.
The \POWHEG $h_\text{damp}$ parameter controls the damping of emissions with high transverse momentum (\pt) in the matrix element calculations and regulates the matching to the \PYTHIA parton shower. Therefore, it controls the amount of additional hard radiation in \ttbar events. 
Since the analysis targets final states with high jet multiplicities and multiple heavy-flavor jets, the modeling of this radiation can affect the event acceptance and the shapes of the discriminating variables. 
The corresponding modeling uncertainty is evaluated by comparing the nominal simulation to dedicated \ttbar samples generated with the $h_\text{damp}$ parameter varied up and down~\cite{Jezo:2018yaf}.
The \ttbar events are further categorized based on the presence of additional heavy-flavor hadrons.
This classification is performed using generator-level information by identifying heavy-flavor hadrons and associating them with their originating partons.
This categorization distinguishes \ttbar events with additional heavy-flavor production beyond the two bottom quarks originating from the top quark decays, defining two categories: $\ttbar+\bbbar$ events, which have additional \bbbar production at the generator level, and $\ttbar+\textrm{non-}\bbbar$ events, which include all remaining \ttbar events.

Additional background processes are grouped separately.
The ``Other \PQt'' category includes single $\PQt$ production, as well as \ttW, \ttZ, and \ttH processes.
The EW category comprises Drell--Yan, \Wjets, and vector boson pair production events.
The remaining background is from QCD multijet events.

\section{Object reconstruction and identification}
\label{sec:Object}
The particle-flow (PF) algorithm~\cite{CMS:2017yfk} aims to reconstruct and identify each individual particle in an event, with an optimized combination of information from the various elements of the CMS detector.
The energy of photons is obtained from the ECAL measurement.
The energy of an electron is determined from a combination of the electron momentum at the primary interaction vertex, as measured by the tracker, the energy of the corresponding ECAL cluster, and the energy sum of all bremsstrahlung photons spatially compatible with originating from the electron track.
The energy of a muon is obtained from the curvature of the corresponding track.
The energy of a charged hadron is determined from a combination of the momentum measured in the tracker and the matching ECAL and HCAL energy deposits, corrected for the response function of the calorimeters to hadronic showers.
Finally, the energy of a neutral hadron is obtained from the corresponding corrected ECAL and HCAL energies.
This analysis uses electrons, muons, jets, and \ptvecmiss reconstructed using the PF algorithm.

The primary vertex (PV) is defined as the vertex corresponding to the hardest scattering in the event, evaluated using tracking information alone, as described in Ref.~\cite{CMS-TDR-15-02}.
The PV is reconstructed from all tracks in the event that are compatible with the LHC beam spot in the transverse plane.
Events are required to have at least one PV within the luminous region, defined by a position along the beamline axis of less than 24.0\cm and a radial distance from the beamline in the transverse plane of less than 2.0\cm, and with the number of degrees of freedom greater than 4.
Additionally, events are required to pass a set of quality filters.

The \ptvecmiss is defined as the negative vector \pt sum of all PF candidates in an event, and its magnitude is denoted by \ptmiss~\cite{CMS:2019ctu}.
The \ptmiss is corrected to account for the jet energy scale corrections applied to the reconstructed jets in the event.

Electron selection is performed using a multivariate analysis (MVA) identification algorithm based on electromagnetic cluster shape information~\cite{electron}. 
Two standard MVA working points are defined, \textsc{loose} and \textsc{tight}, corresponding to identification efficiencies of approximately 90\% and 80\%, respectively. 
The  \textsc{loose} electrons must have $\pt > 15\GeV$, $\abs{\eta} < 2.5$, and a mini-isolation smaller than 0.4 to ensure lepton isolation.
Mini-isolation, used to distinguish leptons from nearby jets, is computed as the inverse ratio of the lepton \pt to the scalar \pt sum of PF candidates within a $\pt$-dependent cone around the lepton~\cite{CMS:2016krz}. 
The \textsc{tight} electrons must have $\pt > 20\GeV$, $\abs{\eta} < 2.5$, and a mini-isolation smaller than 0.1.
Electrons located within the ECAL barrel-endcap transition region ($1.44 < \abs{\eta} < 1.57$) are vetoed in both the \textsc{loose} and \textsc{tight} cases.

Muon selection is based on a cut-based identification algorithm, with \textsc{loose} and \textsc{tight} working points~\cite{MuonID}.
The corresponding identification efficiencies are nearly 100 and 97\%, respectively.
The \textsc{loose} muons must have $\pt > 15\GeV$, $\abs{\eta} < 2.4$, and a mini-isolation smaller than 0.4. 
The \textsc{tight} muons must have $\pt > 20\GeV$, $\abs{\eta} < 2.4$, and a mini-isolation smaller than 0.1, and must satisfy requirements on the transverse (longitudinal) impact parameters less than 0.2 (0.5)\cm with respect to the PV. 
To account for the efficiency differences between data and simulation, scale factors are applied to simulated events for the single-lepton triggers, lepton identification, and mini-isolation.

For each event, hadronic jets are clustered from the reconstructed particles using the infrared- and collinear-safe anti-\kt algorithm~\cite{Cacciari:2008gp,Cacciari:2011ma} with distance parameters of 0.4 and 0.8, referred to as AK4 and AK8 jets, respectively.
The jet momentum is determined as the vector sum of all particle momenta in the jet, and is found from simulation to be, on average, within 5--10\% of the true momentum over the full \pt spectrum and detector acceptance. 
Pileup interactions can contribute extra tracks and calorimetric energy deposits to the jet momentum. To mitigate this effect, tracks identified as originating from pileup vertices are discarded and an offset correction is applied to correct for remaining contributions using the charged-hadron subtraction algorithm~\cite{Sirunyan:2020foa}. These corrections are applied to AK4 jets. 

For AK8 jets, the pileup per particle identification algorithm~\cite{Sirunyan:2020foa,Bertolini:2014bba} is used to mitigate the effect of pileup at the reconstructed-particle level, making use of local shape information, event pileup properties, and tracking information. A local shape variable is defined that distinguishes between collinear and soft diffuse distributions of other particles surrounding the particle under consideration.
The former is attributed to particles originating from the hard scatter and the latter to particles originating from pileup interactions.
Charged particles identified as originating from pileup vertices are discarded.
For each neutral particle, a local shape variable is computed using surrounding charged particles compatible with the primary vertex within the tracker acceptance ($\abs{\eta} < 2.5$), and using both charged and neutral particles outside the tracker acceptance.
The momenta of the neutral particles are then rescaled according to their probability of originating from the PV, as deduced from the local shape variable, superseding the need for jet-based pileup corrections~\cite{Sirunyan:2020foa}.

Jet energy corrections are derived from simulation to bring the measured response of jets to that of particle-level jets on average.
In situ measurements of the momentum balance in dijet, photon + jet, \Zjets, and multijet events are used to account for any residual differences in the jet energy scale between data and simulation~\cite{CMS:2016lmd}.
The jet energy resolution amounts typically to 15--20\% at 30\GeV, 10\% at 100\GeV, and 5\% at 1\TeV~\cite{CMS:2016lmd}.
Additional selection criteria are applied to each jet to remove jets potentially dominated by anomalous contributions from various subdetector components or reconstruction failures~\cite{CMS:2016lmd}.

To avoid double counting of reconstructed objects, for any jet containing an isolated PF lepton among its constituents, the lepton's four-momentum is subtracted from the jet's four-momentum. 

The AK4 jets are considered if they have $\pt > 25\GeV$ and $\abs{\eta} < 2.4$, while AK8 jets are considered if they have $\pt > 200\GeV$ and $\abs{\eta} < 2.4$.
During the 2018 data-taking period, a power outage affected two sectors of the hadron endcap calorimeter, corresponding to the region $-3.0 < \eta < -1.3$ and the azimuthal region $-1.57 < \phi < -0.87$.
This resulted in degraded jet energy measurements in that region and can lead to spurious \ptmiss in the signal region.
To mitigate this effect, jets within the affected region are excluded from the analysis.

The identification of AK4 jets originating from bottom quarks is performed using the \DeepJet discriminant~\cite{Bols_2020, BTV-16-002, CMS-DP-2023-005} based on a deep neural network.
Jets with a \DeepJet discriminant value exceeding a working point corresponding to an efficiency of 80\% and a mistag rate of 1\% are identified as bottom quark tagged jets (denoted as \PQb jets).
To correct for differences in \PQb tagging performance between data and simulation, scale factors are applied to jets that are matched to particle-level \PQb hadrons.

The \textsc{ParticleNet} multiclass algorithm is designed to distinguish jets originating from top quarks or \PW bosons from those arising from QCD processes~\cite{PN}.
The tagger working points are defined in simulation and calibrated in data using a single-muon control sample enriched in \ttbar events.
The AK8 jets with $\pt > 300\GeV$ and a \textsc{ParticleNet} discriminant value above the threshold corresponding to a mistag rate of approximately 1\% for QCD jets are identified as \PQt tagged. This working point corresponds to an approximate per jet tagging efficiency of 90\% for hadronically decaying top quark jets~\cite{CMS-DP-2025-010}.
Similarly, AK8 jets with $\pt > 200\GeV$ and a discriminant value above the threshold corresponding to a mistag rate of approximately 1\% for QCD jets are identified as \PW tagged, with an approximate per jet tagging efficiency of 60--70\% for hadronically decaying \PW boson jets~\cite{CMS-DP-2025-010}. 
Because the same collection of AK8 jets is used for both \PW and \PQt tagging, jets satisfying both tagging criteria are classified as \PQt tagged. 
The quoted efficiency values are approximate and vary depending on the data-taking period and AK8 jet \pt.

The AK8 jets are used for Lorentz-boosted object reconstruction, while AK4 jets are primarily used for bottom quark jet identification. 
No explicit overlap removal is applied between the AK4 and AK8 jets. Therefore, AK4 jets may geometrically overlap with AK8 jets used for \PW boson and \PQt tagging. The tagging efficiencies and mistag rates depend on the jet transverse momentum, and differences between data and simulation are corrected using scale factors that depend on the jet \pt and the working point.

\section{Event selection}
\label{sec:Selection}
The initial selection requires one \textsc{tight} lepton and no additional \textsc{loose} leptons.
The selected lepton must be either an electron with $\pt > 35\GeV$ or a muon with $\pt > 30\GeV$, satisfying the respective single-lepton trigger.
The trigger efficiencies are studied in data and simulation using a tag-and-probe technique~\cite{CMS:2011aa} based on leptonically decaying \PZ boson events.
Lepton \pt thresholds are chosen above the point where the triggers operate above their full efficiency to ensure that the trigger performs reliably. 
The single-electron triggers require an isolated electron with $\pt > 32\GeV$ and $\abs{\eta} < 2.1$.
The single-muon triggers require an isolated muon with $\pt > 24\GeV$ for the non-2017 data set, or $\pt > 27\GeV$ for the 2017 data set, and $\abs{\eta} < 2.5$.
Discrepancies in the trigger efficiency between data and simulation are corrected using scale factors that depend on \pt and $\eta$.

\begin{figure*}[hbtp]
\centering
\includegraphics[width=0.7\textwidth]{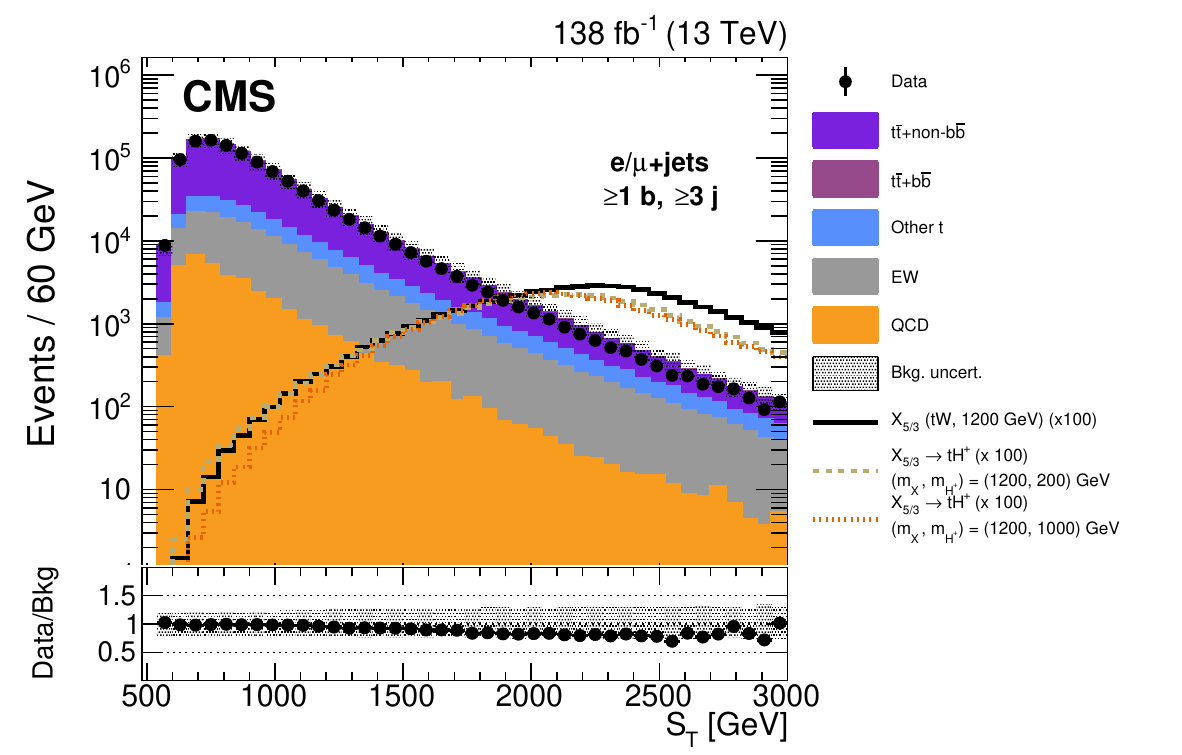}
\caption{The distribution of \ST after the initial selection for data and simulated background processes.
	The background uncertainty includes both statistical and systematic components.}
\label{fig:ST}
\end{figure*}

The initial selection requires events to have at least three AK4 jets, each with $\pt > 30\GeV$ and $\abs{\eta} < 2.4$.
At least one of the AK4 jets must be \PQb tagged, and the event must satisfy $\ptmiss > 30\GeV$ to suppress multijet background.
Additionally, the scalar \pt sum of all reconstructed AK4 jets with $\pt > 30\GeV$ and $\abs{\eta} < 2.4$, denoted by \HT, must exceed 500\GeV. 
Any AK8 jet must have $\pt > 200\GeV$ and $\abs{\eta} < 2.4$. 
The distribution of \ST, defined as the scalar sum of the lepton \pt, \ptmiss, and \HT, is shown in Fig.~\ref{fig:ST} for the combined 2016--2018 data set. 
An overprediction of the yield is observed at high \ST, primarily due to mismodeling of \ttbar production, which is accounted for by the corresponding systematic uncertainties. 
The signal is characterized by larger \ST values than the background processes.
This observable provides strong discriminating power and is a key component of the analysis strategy.

\begin{figure*}[hbtp]
\centering
\includegraphics[width=0.7\textwidth]{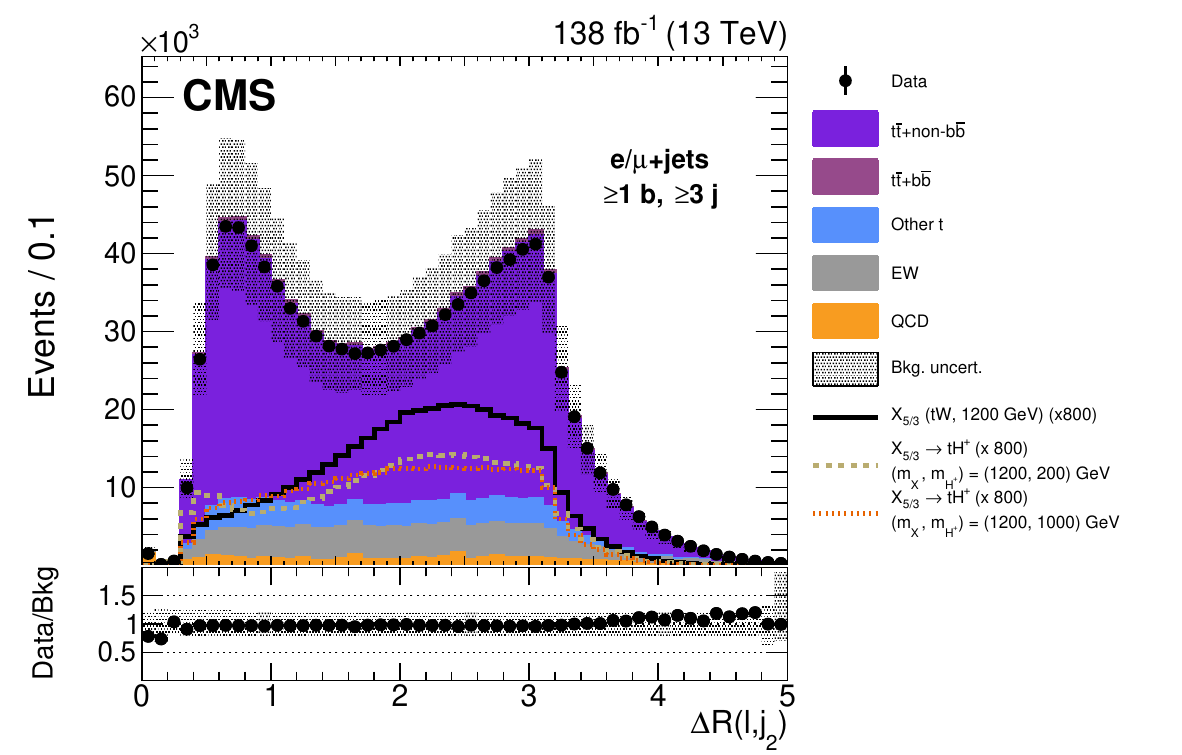}
\caption{The distribution of the angular separation between the lepton $\Pell$ and the jet subleading in \pt after the initial selection for data and simulated background processes.
	The background uncertainty includes both statistical and systematic components.}
\label{fig:deltaRjet2}
\end{figure*}

In addition to the initial selection, the angular separation, $\DR = \sqrt{\smash[b]{(\Delta\eta)^2 + (\Delta\phi)^2}}$, between the selected lepton and the subleading AK4 jet in \pt, denoted by $\text{j}_2$, is required to satisfy $\Delta R(\Pell, \text{j}_2) > 1.0$ in the final event selection. 
The variable probes the event topology through the spatial separation of the lepton and the subleading jet. 
The $\Delta R(\Pell, \text{j}_2)$ distribution after the initial selection is shown in Fig.~\ref{fig:deltaRjet2} for the combined 2016--2018 data set. 
The distribution exhibits a bimodal structure, with signal events preferentially populating the region at larger $\Delta R(\Pell, \text{j}_2)$ values. The requirement $\Delta R(\Pell, \text{j}_2) > 1.0$ retains approximately 90\% of the signal in the conventional decay mode and 84--89\% in the exotic decay mode, depending on the \PHpm mass. 
The requirement was determined through optimization studies and was found to improve the expected sensitivity of the analysis by preserving a high signal efficiency.  

A signal-depleted control region is defined by inverting this requirement, $\Delta R(\Pell, \text{j}_2) < 1.0$, while keeping all other selection criteria unchanged.
Signal contamination in this region is found to be 16\% or less for the 1200\GeV $\PX_{5/3}$ signal mass point.
This control region is used for validation purposes only and is not included in the signal extraction.

To enhance signal-to-background separation, a multivariate discriminant based on \textsc{XGBoost}, a gradient boosting library~\cite{Chen:2016}, is employed.
A boosted decision tree (BDT) is trained using simulated events for $\PX_{5/3}$ signals with masses of 1200\GeV in the exotic decay mode and 1300\GeV in the conventional decay mode, together with \ttbar background events, after applying the final selection criteria. 
The BDT is trained separately for the conventional and exotic decay modes, and independently for each data set corresponding to a different data-taking period, to account for variations in detector conditions and data quality that may affect the performance of the discriminant. 
The BDT provides effective discrimination for other $\PX_{5/3}$ signal mass points not used during training, with the area under the receiver operating characteristic curve reaching at least 0.9 for masses above 1000\GeV. 
Separate \textsc{XGBoost} outputs are used for each \PHpm mass point to enable independent statistical interpretations, such as limit setting, for each mass hypothesis. 
The simulated samples are divided equally into training, validation, and application subsets.
The \textsc{XGBoost} \emph{gain} metric ranks features based on their contribution to improving the model performance when used to split decision tree nodes.
Only the relative ordering of variables is used for the input selection, and no quantitative interpretation of the gain values is made.
The 20 highest-ranked variables, listed in Tables~\ref{tab:bdtX53tW} and~\ref{tab:bdtX53tH}, are used as inputs to the multivariate discriminant in the conventional and exotic decay modes, respectively. 
Restricting the input to the 20 highest-ranked variables balances discrimination power with the ability to validate the input variable modeling, while avoiding the inclusion of variables with little additional discrimination power. 
The input variables are validated through studies of data-to-simulation agreement and checks of intervariable correlations.
In the conventional decay mode, variables related to \PW bosons and \PW-tagged jets are ranked most highly, while in the exotic decay mode, variables associated with bottom quarks and \PQb-tagged jets are dominant.

All composite input variables used in the BDT are constructed from reconstructed objects passing the final event selection.
Unless otherwise stated, jets refer to AK4 jets with $\pt > 30\GeV$ and $\abs{\eta} < 2.4$.
The sphericity is defined from the eigenvalues of the momentum tensor constructed from all jets in the event~\cite{Fox:1978vu}.
The variable denoted as the invariant mass of all jets and leptonically decaying \PW bosons is defined as the four-vector sum of all jets and the reconstructed \PW boson candidate. The reconstructed \PW boson four-momentum is obtained by summing the lepton and neutrino four-momenta, where the neutrino longitudinal momentum is determined using the \PW boson mass constraint. 
Variables referring to the $n$th-leading jet or \pt correspond to the ordering of selected AK4 jets by \pt. 
The quantity \HTb is defined as the scalar \pt sum of all \PQb-tagged AK4 jets. 
The $\HT/\HTj$ variable is defined as the ratio of \HT to \HTj, where \HTj denotes the scalar \pt sum of the three highest \pt AK4 jets.

\begin{table*}[h!tb]
\centering
\topcaption{Ranking of input variables for the BDT trained on the 2017 data set using signal samples in the conventional decay mode with an $\PX_{5/3}$ mass of 1300\GeV.
	The ranking is based on the gain metric from \textsc{XGBoost}.}
\begin{scotch}{ll}
Rank & Variable \\
\hline
1 & \ST \\
2 & Number of \PW-tagged AK8 jets \\
3 & Invariant mass of all jets and leptonically decaying \PW bosons \\
4 & Invariant mass of the $\Pell$ and angularly closest \PQb jet pair \\
5 & Number of \PQt-tagged AK8 jets \\
6 & Invariant mass of the $\Pell$ and jet with subleading \pt \\
7 & Invariant mass of the $\Pell$ and jet with third-highest \pt \\
8 & Invariant mass of the $\Pell$ and leading \PQb jet pair \\
9 & Sphericity \\
10 & Minimum invariant mass of the $\Pell$ and any \PQb jet pair \\
11 & Minimum angular distance between the $\Pell$ and any AK4 jet \\
12 & Transverse mass of the $\Pell$ and \ptvecmiss \\
13 & \ptmiss \\
14 & \pt of the system of the two leading \PQb jets and \ptvecmiss \\
15 & Minimum angular distance between the $\Pell$ and any \PQb jet pair \\
16 & Lepton momentum \\
17 & Third-highest \DeepJet discriminant value \\
18 & Second-highest \DeepJet discriminant value \\
19 & \DR between the $\Pell$ and the \PQb jet yielding the minimum invariant mass \\
20 & \DR between the $\Pell$ and the jet yielding the minimum invariant mass \\
\end{scotch}
\label{tab:bdtX53tW}
\end{table*}

\begin{table*}[h!tb]
\centering
\topcaption{Ranking of input variables for the BDT trained on the 2017 data set using signal samples in the exotic decay mode with an $\PX_{5/3}$ mass of 1200\GeV and an $\PHpm$ mass of 200\GeV.
	The ranking is based on the gain metric from \textsc{XGBoost}.}
\begin{scotch}{ll}
Rank & Variable \\
\hline
1 & \ST \\
2 & Third-highest \DeepJet discriminant value \\
3 & Sixth-leading jet \pt \\
4 & Number of \PQt-tagged AK8 jets \\
5 & Invariant mass of all jets and leptonically decaying \PW bosons \\
6 & Sphericity \\
7 & Invariant mass of the $\Pell$ and leading \PQb jet pair \\
8 & \HTb \\
9 & Invariant mass of the $\Pell$ and angularly closest \PQb jet pair \\
10 & Second-highest \DeepJet discriminant value \\
11 & \DeepJet discriminant value of the jet with the highest \pt \\
12 & Transverse mass of the $\Pell$ and \ptvecmiss \\
13 & Fifth-leading jet \pt \\
14 & Angular distance between the $\Pell$ and angularly closest \PQb jet pair \\
15 & Invariant mass of the $\Pell$ and the jet with third-highest \pt \\
16 & Minimum invariant mass of the $\Pell$ and any \PQb jet pair \\
17 & \ptmiss \\
18 & \HT/\HTj \\
19 & \DeepJet discriminant value of the jet with second-highest \pt \\
20 & Invariant mass of the $\Pell$ and the jet with the second-highest \pt \\
\end{scotch}
\label{tab:bdtX53tH}
\end{table*}

Signal selection efficiencies for the conventional and exotic decay modes are shown in Table~\ref{tab:eff_x53} for the lowest and highest $\PX_{5/3}$ mass points. 
The efficiencies decrease at higher $\PX_{5/3}$ masses because the objects in the final state are more Lorentz-boosted and therefore more collimated.
The efficiencies are lower for the $\PX_{5/3} \to \PQt\PSHp$ decay mode because of the more complex final states, including the presence of six bottom quarks. 
For heavier charged Higgs bosons, the decay products are more Lorentz-boosted, leading to higher efficiencies because of the increased likelihood of satisfying the final selection thresholds.

\begin{table}[htbp]
\centering
\topcaption{Signal selection efficiencies for the conventional and exotic decay modes, defined as the ratio of the number of signal events passing the final event selection to the total number of generated events in the simulated signal sample. The quoted efficiencies include the branching fraction into the single-lepton final state. The low and high masses correspond to 700 and 1700\GeV in the conventional decay mode or 600 and 1900\GeV in the exotic decay mode. The mass values in parentheses refer to the \PHpm mass. The listed uncertainties are statistical only.}
\begin{scotch}{cc}
	Low mass (\%) & High mass (\%) \\
\hline
\multicolumn{2}{l}{$\PX_{5/3}\to\PQt\PW$} \\
$30.73 \pm 0.05$ & $29.77 \pm 0.05$ \\[\cmsTabSkip]
\multicolumn{2}{l}{$\PX_{5/3}\to\PQt\PSHp\,(200\GeV)$} \\
$22.94 \pm 0.04$ & $17.67 \pm 0.04$ \\[\cmsTabSkip]
\multicolumn{2}{l}{$\PX_{5/3}\to\PQt\PSHp\,(1000\GeV)$} \\
\NA & $23.93 \pm 0.04$ \\
\end{scotch}
\label{tab:eff_x53}
\end{table}

The low-BDT-score region is excluded by requiring a BDT score greater than 0.1 to reduce potential biases from regions with limited discriminating power.
This region is strongly dominated by background and provides negligible sensitivity to the signal hypothesis, but the large number of events in this region could cause the results to be driven by any data-to-simulation differences in the BDT tail.
The exclusion of this region improves the stability of the statistical interpretation with negligible impact on the sensitivity. The signal is extracted from these \textsc{XGBoost} distributions. 

\section{Systematic uncertainties}
\label{sec:Systematics}

Systematic uncertainties, listed in Table~\ref{tab:sys-error}, are considered from several sources that affect the normalization and shape of the background or signal distributions.

\begin{table*}[h!tb]
\centering
\topcaption{Summary of the systematic uncertainties considered. 
	Uncertainties labeled as ``Shape'' correspond to variations whose templates are normalized to the total yield and therefore affect only the distribution shape. Uncertainties labeled as ``Correlated'' are correlated across years. Here, SD stands for a standard deviation. Uncertainties are symmetric.}
\begin{scotch}{l c c c c}
Source & Uncertainty & Signal & Background & Correlated \\
\hline
Normalization only & & & & \\
Integrated luminosity 2016--2018 & 0.73\% & Yes & All & Partially \\
Electron identification & 1.5\% & Yes & All & Yes \\
Electron isolation & 2.5\% & Yes & All & Yes \\
Electron trigger & 5\% & Yes & All & Yes \\
Muon identification & 1\% & Yes & All & Yes \\
Muon isolation & 2.5\% & Yes & All & Yes \\
Muon trigger & 5\% & Yes & All & Yes \\
Cross section \ttbar+\xspace\bbbar & 10\% & No & \ttbar+\xspace\bbbar & Yes \\
Cross section \ttbar+\xspace\textrm{non-}\bbbar & 10\% & No & \ttbar+\xspace\textrm{non-}\bbbar & Yes \\
\ttH cross section~\cite{Czakon:2011xx} & 20\% & No & \ttH & Yes \\
Single top cross section~\cite{Aliev_2011, Kant_2015} & 4\% & No & Single $\PQt$ & Yes \\
EW cross section~\cite{Campbell_2005} & 3.8\% & No & EW & Yes \\
\ttbar sample $h_\text{damp}$ & 1 SD & No & \ttbar & Yes \\[\cmsTabSkip]
Shape and normalization & & & & \\
Prefiring & 1 SD & Yes & All & No \\
Pileup & 4.6\% & Yes & All & Yes \\
Jet pileup & 1 SD(\pt) & Yes & All & Yes \\
Jet energy scale & 1 SD(\pt,$\eta$) & Yes & All & No \\
Jet energy resolution & 1 SD($\eta$) & Yes & All & No \\
Top quark tagging & 1 SD($\eta$) & Yes & All & No \\
\PW tagging & 1 SD ($\eta$) & Yes & All & No \\
\DeepJet correlated & 1 SD(\pt) & Yes & All & Partially \\
Top quark \pt reweighting & 1 SD(\pt) & No & \ttbar & Yes \\
PDF & 1 SD & Shape & All & Yes \\
ISR & 1 SD & Yes & All & Yes \\
FSR & 1 SD & Yes & All & Yes \\
$\mu_\mathrm{F}$ & 1 SD & Shape & All & Yes \\
$\mu_\mathrm{R}$ & 1 SD & Shape & All & Yes \\
MC statistical & & Yes & All & No \\
\end{scotch}
\label{tab:sys-error}
\end{table*}

The integrated luminosities for the 2016, 2017, and 2018 data-taking periods have both uncorrelated uncertainties and correlated components across data-taking years, varying from 0.04\% to 1.18\%~\cite{CMS-LUM-17-003, CMS-PAS-LUM-20-001}.
The uncertainty in the integrated luminosity for the combined 2016--2018 data set is 1.6\%.
The uncertainties in lepton reconstruction, isolation, and trigger efficiency are treated as flat rates, with the reconstruction uncertainties derived from the \pt- and $\eta$-dependent variations, the isolation uncertainties are centrally provided, and a 5\% uncertainty is assigned to the trigger efficiency.

During the 2016 and 2017 data-taking periods, a gradual timing shift in the inputs to the ECAL L1 trigger in the $\abs{\eta} > 2.0$ region led to a trigger inefficiency referred to as ``prefiring''~\cite{CMS:2020cmk}.
This effect caused the trigger to miss between 10--20\% of events containing an electromagnetic object with $\pt \gtrsim 50\GeV$ or a jet with $\pt \gtrsim 100\GeV$ in the $\abs{\eta}$ range of 2.5--3.0, depending on the \pt and $\eta$ of the trigger object.
Correction factors derived from data are applied to simulation-based acceptances. 
The effect is negligible for this analysis, as the affected region lies outside the trigger acceptance and the jet $\eta$ requirement.

The pileup reweighting uncertainty is evaluated by varying the total inelastic cross section of 69.2\unit{mb} by 4.6\%~\cite{CMS:2018mlc}. 
Jet energy scale uncertainties are assessed by varying the jet energy corrections by one standard deviation in the simulation, followed by reprocessing the full analysis to derive new discriminant shapes.
Jet energy resolution uncertainties are evaluated similarly.
To account for discrepancies in the modeling of the top quark \pt spectrum, a correction is applied to the generated top quark \pt in \ttbar events~\cite{CMS:2018nzx}.
The uncertainty from top quark \pt reweighting is estimated by producing an alternative template without applying the reweighting. The difference between the nominal weighted and the unweighted distributions is taken as a one-sided systematic variation, which is then symmetrized about the nominal distribution to define upward and downward variations. 
The uncertainty in the jet pileup identification scale factor is applied as a function of the jet \pt and $\eta$.

The PDF uncertainties are evaluated using 100 NNPDF3.1 replicas, each implemented using PDF weights.
The variations in each bin are ordered, and the 16\% and 84\% quantiles are taken to define the one standard deviation down and up shifts to account for PDF uncertainties~\cite{NNPDF:2017mvq}. 
Uncertainties due to missing higher-order corrections are obtained by varying the renormalization ($\mu_\mathrm{R}$) and factorization scales ($\mu_\mathrm{F}$) up and down in the simulation. 
For the PDF and $\mu_\mathrm{R}$ and $\mu_\mathrm{F}$ scale uncertainties, the varied templates in signal are normalized to the total yield so that these uncertainties only affect the signal shape.
The initial-state radiation (ISR) and final-state radiation (FSR) uncertainties are evaluated using parton shower variation weights provided in the simulated samples when available. 
These variations affect both the normalization and shape of the corresponding simulated processes. 
Additional production cross section uncertainties are applied for background processes, independent of the variations described above. 
The robustness of the analysis with respect to the assumed 10\% \ttbar+\xspace\bbbar and \ttbar+\xspace\textrm{non-}\bbbar normalization uncertainties was verified by varying their magnitudes up to 20\%, yielding a negligible impact on the expected sensitivity. 
An uncertainty associated with the modeling of hard radiation in \ttbar events is estimated by comparing the nominal \ttbar simulation with dedicated samples generated using the $h_\text{damp}$ up and down variations, and the resulting differences are taken as the systematic uncertainty.

The various jet-tagging algorithms also have corresponding uncertainty sources.
Uncertainties in the \PQb tagging efficiencies are calculated using the standard \DeepJet calibration uncertainty components~\cite{CMS-DP-2023-005}.
Sources originating from the limited statistical precision of the calibration samples are treated as uncorrelated across data-taking years, while all other \PQb tagging uncertainty components are taken to be fully correlated across years.

Statistical uncertainties arising from the finite size of the simulated samples are also included.
To reduce bin-by-bin fluctuations in the systematically varied templates, a smoothing procedure based on the locally weighted scatterplot smoothing (LOWESS) algorithm is applied~\cite{Cleveland01091988}, using a smoothing parameter of 0.8. 
The smoothing is performed on the bin-by-bin ratios of the systematically varied templates to the nominal template, and the resulting smooth ratios are applied to the nominal distribution. 
The effect of the smoothing procedure is validated by inspecting the up and down variations of all affected systematic uncertainties before and after smoothing to ensure that the procedure reduces statistical fluctuations while preserving the overall shape and normalization of the variations. 
The relative impact of the statistical and systematic uncertainties on the sensitivity of the analysis is discussed in Section~\ref{sec:Statistics}.

\section{Results}
\label{sec:Statistics}

Upper limits on the $\PX_{5/3}$ pair production cross section are set using a binned likelihood fit to the distributions of the \textsc{XGBoost} output in the signal region.
The fit is performed simultaneously across all analysis channels, corresponding to the lepton flavors and data-taking years.

To mitigate the impact of statistical fluctuations, the BDT output templates are rebinned, starting from the highest BDT score bin and successively merging adjacent bins until the statistical uncertainty in the total background yield in each bin is below 20\% (30\%) for the conventional (exotic) decay mode. 
The different thresholds reflect the lower event yield in the exotic decay mode and were chosen based on studies of the statistical stability of the fit.  
A common binning scheme, determined from the statistically most limited data-taking period (2016), is applied uniformly across all data-taking periods.

The systematic uncertainties listed in Table~\ref{tab:sys-error} are treated as nuisance parameters.
The nuisance parameters corresponding to each systematic uncertainty are assigned log-normal constraints for normalization-only uncertainties and Gaussian constraints for shape uncertainties, while the signal cross section is unconstrained.
The \ttbar background is modeled using separate templates for the \ttbar+\xspace\bbbar and \ttbar events without additional \bbbar production, and their normalizations are constrained by the uncertainties on their respective production cross section.

\begin{figure*}[hbtp]
\centering
\includegraphics[width=0.7\textwidth]{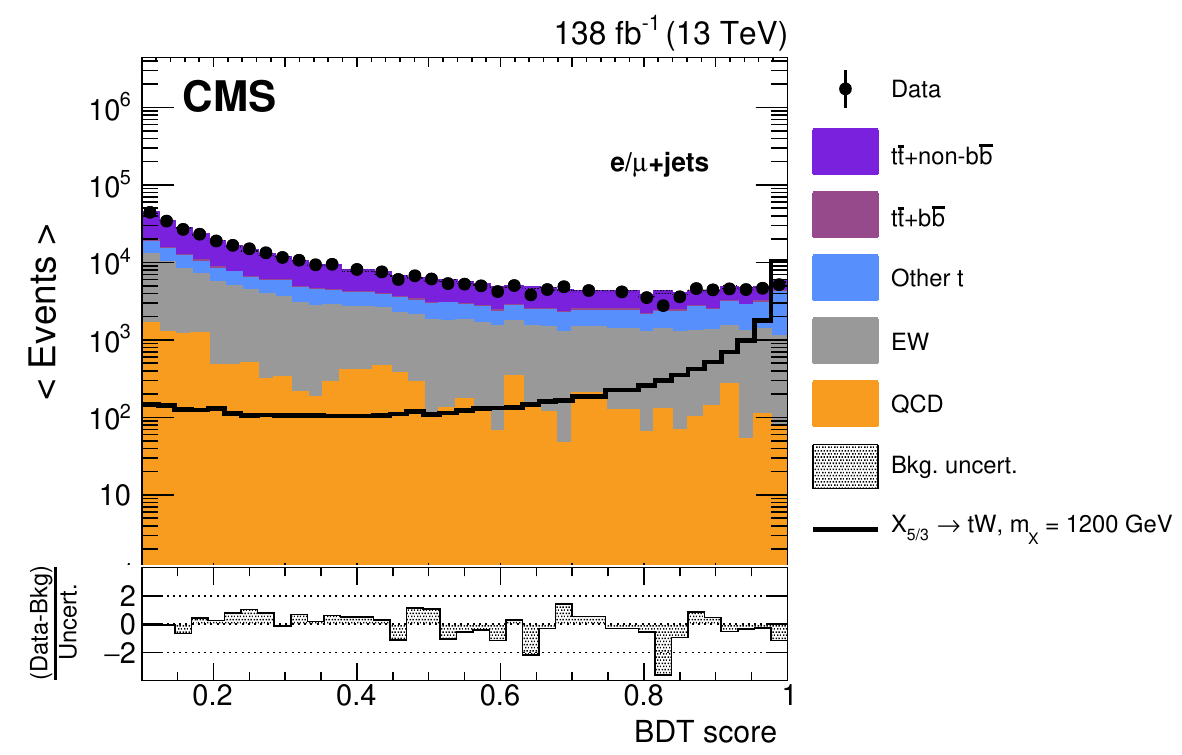}
\caption{Post-fit distribution of the \textsc{XGBoost} output for the $\PX_{5/3} \to \PQt\PW$ model in data and simulation in the signal region after rebinning. The background contributions are shown after the fit to data, while the signal distribution is shown using the pre-fit prediction, normalized to the theoretical cross section of 0.0118\unit{pb}. The distributions are normalized by the bin width to account for the variable bin sizes. The lower panel shows the difference between the data and the post-fit background prediction, divided by the total post-fit uncertainty in the background prediction.}\label{fig:XGB1300rebinned}
\end{figure*}

\begin{figure*}[hbtp]
\centering
	\includegraphics[height=0.36\textwidth]{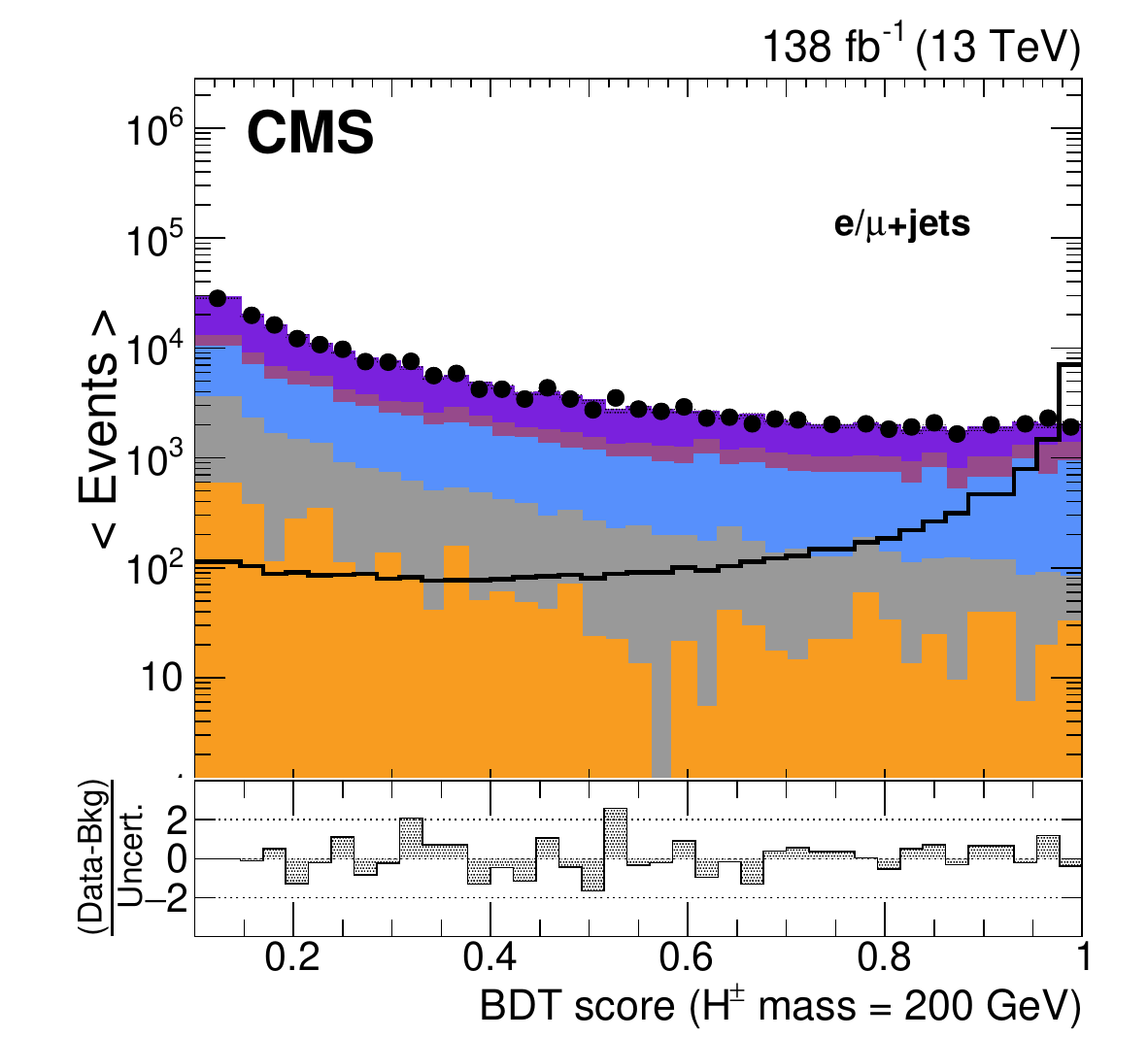}
	\includegraphics[height=0.36\textwidth]{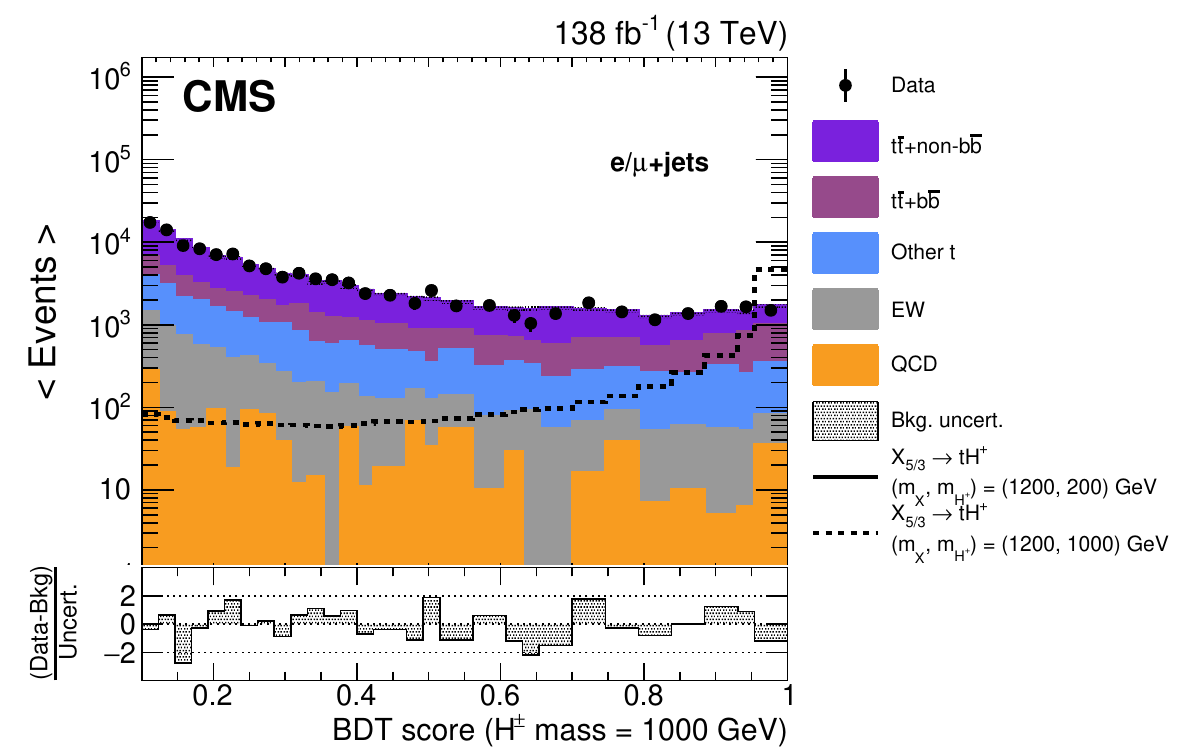}
	\caption{Post-fit distributions of the \textsc{XGBoost} outputs for the $\PX_{5/3} \to \PQt\PSHp$ model with the lowest (left) and highest (right) $\PHpm$ mass hypotheses, 200 and 1000\GeV, respectively, in data and simulation in the signal region after rebinning. Notations as in Fig.~\ref{fig:XGB1300rebinned}.}	
\label{fig:XGBmHrebinned}
\end{figure*}

The resulting post-fit distributions of the BDT outputs in the signal region, after the rebinning procedure, are shown combined across all data-taking years.
The distribution for the conventional decay mode is shown in Fig.~\ref{fig:XGB1300rebinned}, while those for the exotic decay mode are shown in Fig.~\ref{fig:XGBmHrebinned}. 
The residual mismodeling in the \ttbar background, including the known overprediction at high \ST, is constrained by the simultaneous fit to data through the relevant nuisance parameters, resulting in good agreement between data and prediction after the fit.

No significant excess of data over the SM expectation is observed.
Upper limits are set at 95\% \CL on the production cross section, computed using a profile likelihood test statistic in the asymptotic approximation~\cite{Cowan:2010js} and the $\CLs$ criterion~\cite{CLS2,CLS1}.
The results are obtained using the CMS statistical analysis tool \textsc{Combine}~\cite{CMS:2024onh}, which is based on the \textsc{RooFit}~\cite{Verkerke:2003ir} and \textsc{RooStats}~\cite{Moneta:2010pm} frameworks.

Several statistical tests are conducted in the signal region, including saturated likelihood-based goodness-of-fit tests that assess the agreement between the SM background prediction and the data.
The goodness-of-fit tests yield $p$-values in the range 0.32--0.78, indicating good agreement.
The nuisance parameters and their uncertainties are compared before and after the fit.
The impact of each nuisance parameter is evaluated by observing the variation in the signal strength when the nuisance parameter is varied by one standard deviation.
The control region defined in the analysis is used for validation purposes and is subjected to the same statistical tests before the final likelihood fit is performed. Good agreement between the data and the background prediction is also observed.

The sensitivity of the analysis is driven by the high-score BDT region.
The analysis is primarily limited by systematic uncertainties, while statistical uncertainties have a negligible impact ($<$1\%) on the expected limits. 
The dominant systematic uncertainties in the conventional decay mode arise from $\mu_{\mathrm{R}}$ and $\mu_{\mathrm{F}}$ scales and ISR modeling, corresponding to variations in the signal yield up to 16\% for the 1200\GeV $\PX_{5/3}$ mass point. 
In the exotic decay mode, the dominant systematic uncertainties are associated with the \ttbar background normalization, FSR and ISR modeling, and the $h_\text{damp}$ parameter choice, corresponding to variations in the signal yield of up to 15\% for the 1200\GeV $\PX_{5/3}$ mass point. 

\subsection{Conventional decay mode}

\begin{figure*}[hbtp]
\centering
\includegraphics[width=0.7\textwidth]{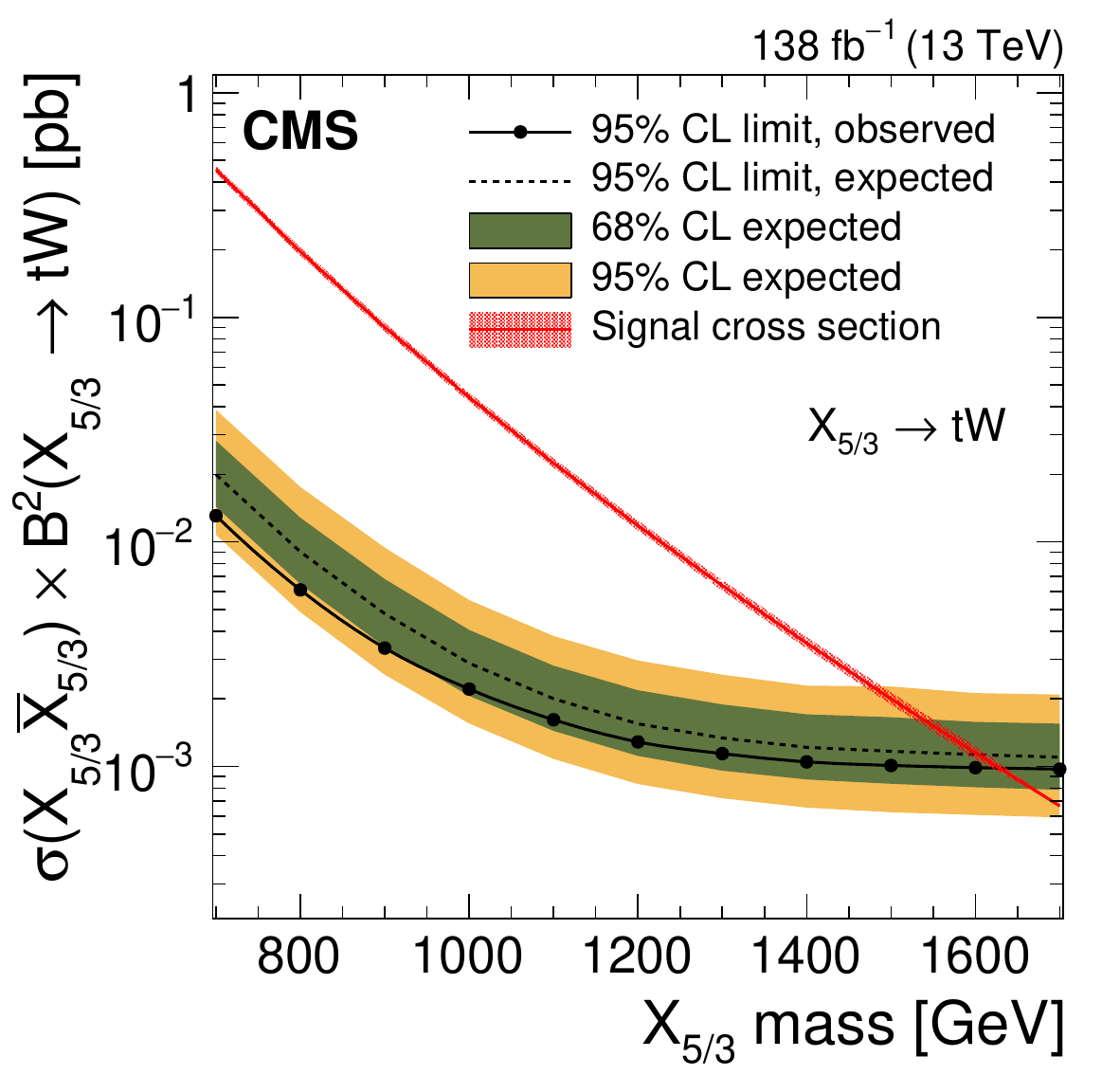}
\caption{The expected and observed upper limits on the $\PX_{5/3}$ pair production cross section for the combined 2016--2018 data set.
	The inner green and outer yellow bands show the one and two standard deviation uncertainty bands, respectively, around the expected limit under the background-only hypothesis.}
\label{fig:LimitsX53comb}
\end{figure*}

For the combined 2016--2018 data set, with an integrated luminosity of 138\fbinv, the expected and observed limits on the $\PX_{5/3}$ cross section are shown in Fig.~\ref{fig:LimitsX53comb}.
Masses of the right-handed $\PX_{5/3}$ in the conventional decay mode below 1600 and 1630\GeV are excluded for the expected and observed limits, respectively.
The inner green and outer yellow bands indicate the regions containing 68 and 95\%, respectively, of the distribution of limits expected under the background-only hypothesis.
The upper limits are compared with the theoretical $\PX_{5/3}$ pair production cross section at NNLO, and the PDF and $\mu_\mathrm{R}$ and $\mu_\mathrm{F}$ scale uncertainties in the theoretical cross section are shown as a band around the theoretical prediction~\cite{X53xsec_paper}.
The observed limit lies below the expected limit and remains within the two standard deviation uncertainty band at low mass.
This behavior is driven by a deficit of events in the highest BDT bins, which provide the strongest discrimination between signal and background.
In the highest BDT bin, the data lie slightly below the post-fit background prediction, particularly in the electron channel, leading to a stronger observed exclusion compared with the SM expectation. 
These are the most restrictive limits on the $\PX_{5/3}$ to date in this decay mode.

\subsection{Exotic decay mode}

\begin{figure}[hbtp]
\centering
\includegraphics[width=0.49\textwidth]{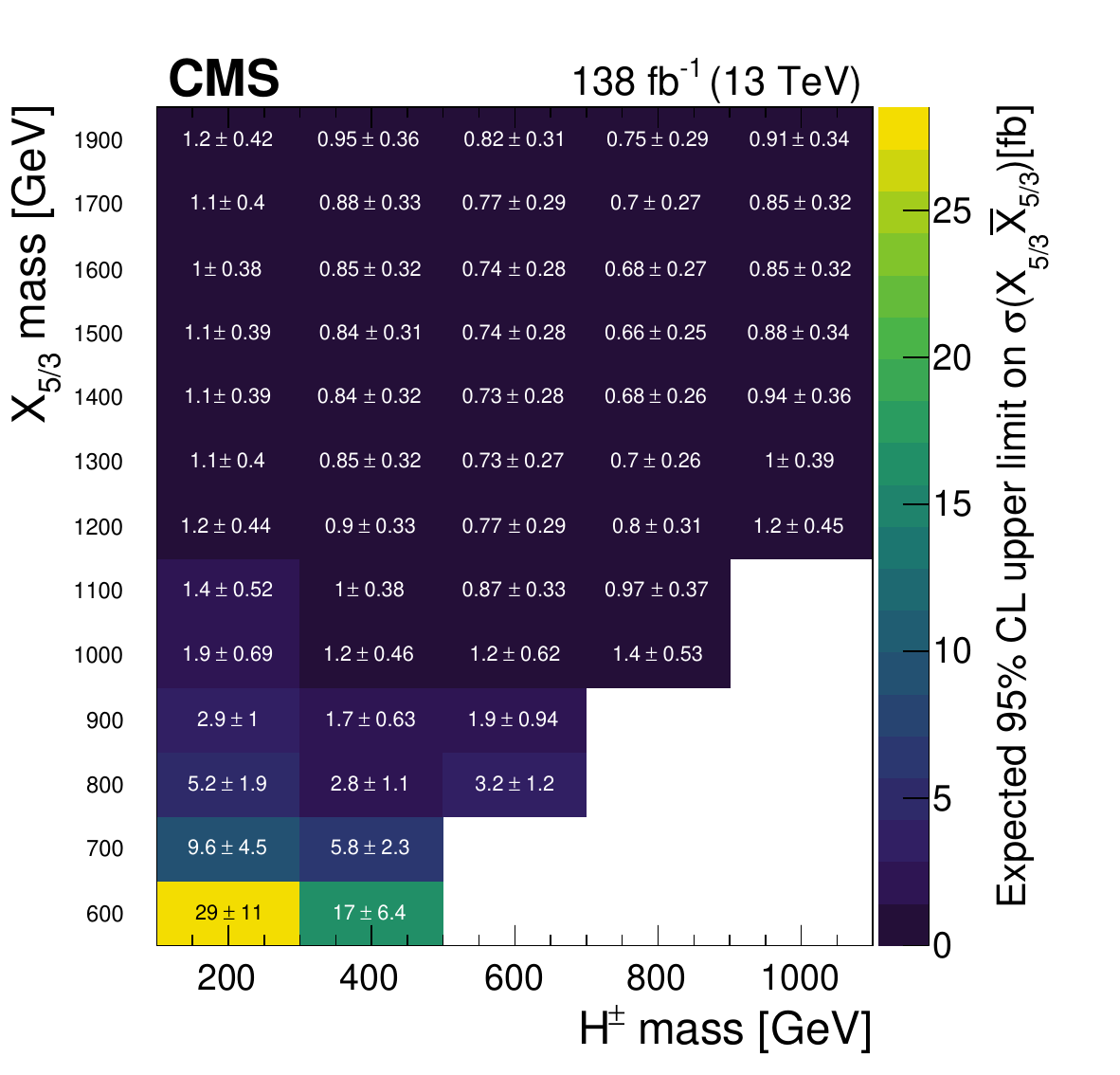}
\includegraphics[width=0.49\textwidth]{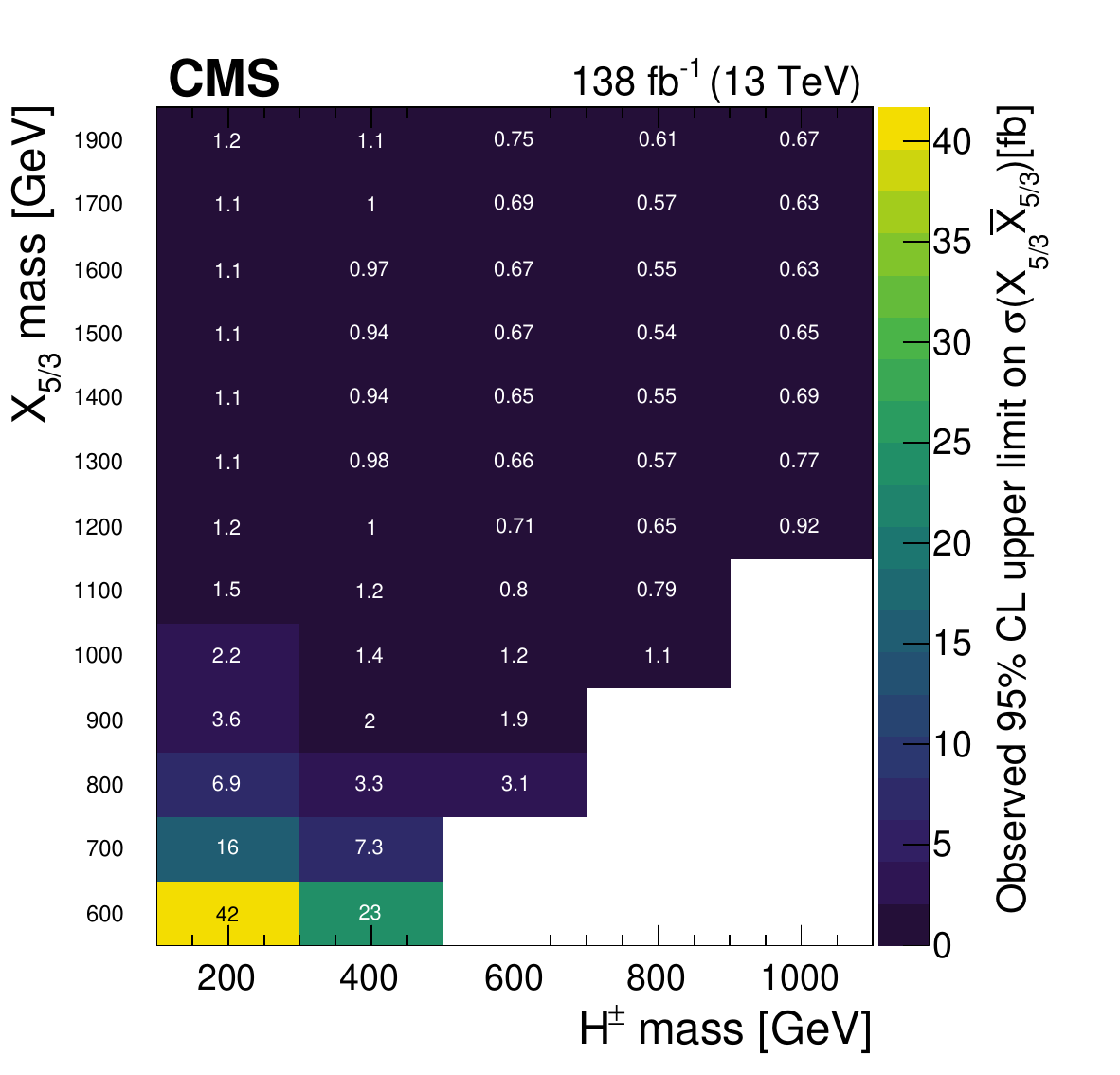}
\caption{The expected (\cmsLeft) and observed (\cmsRight) 95\% \CL upper limits on the production cross section for pair-produced $\PX_{5/3}$ particles in the $\PX_{5/3} \to \PQt\PSHp$ model, shown as functions of the $\PX_{5/3}$ and $\PHpm$ masses, for the combined 2016--2018 data set. 
	The interpretation assumes that the $\PX_{5/3}$ decays exclusively to a top quark and a charged Higgs boson, while the latter decays exclusively into a top and bottom quark pair. 
	For the expected limits, the uncertainties shown correspond to the half-width of the 68\% expected band. 
	The empty cells correspond to mass combinations for which the $\PX_{5/3} \to \PQt\PSHp$ decay is kinematically forbidden. The expected and observed lower limits on the $\PX_{5/3}$ mass are 1620--1690 and 1620--1740\GeV, respectively, for $\PHpm$ masses between 200 and 1000\GeV.}
\label{fig:Limits2D}
\end{figure}

The expected and observed limits on the $\PX_{5/3}$ cross section exclude $\PX_{5/3}$ masses in the exotic decay mode below 1620--1690 and 1620--1740\GeV, respectively, for $\PHpm$ masses between 200--1000\GeV.
A two-dimensional plot showing the expected and observed 95\% \CL upper limits on the production cross section for pair-produced $\PX_{5/3}$ particles is shown in Fig.~\ref{fig:Limits2D}.
The observed limits are generally consistent with the expected limits within the one standard deviation uncertainty band.
The largest deviation occurs at low $\PX_{5/3}$ mass and an \PHpm mass of 200\GeV, but remains within the two standard deviation uncertainty band.
The small deviations arise from statistical fluctuations in the high-BDT-score regions and do not indicate any significant excess or deficit relative to the SM prediction. 
These are the first limits on the $\PX_{5/3}$ in this decay mode.

\section{Summary}
\label{sec:Conclusion}
A search for the pair production of heavy partners of the top quark with an exotic electric charge of 5/3 times the charge of the positron, $\PX_{5/3}$, has been presented.
The $\PX_{5/3}$ is assumed to decay either into a top quark and a \PW boson, or into a top quark and a charged Higgs boson, \PHpm, that subsequently decays into a top and bottom quark pair.
The search uses proton-proton collision data collected during 2016--2018 at a center-of-mass energy of 13\TeV, corresponding to an integrated luminosity of 138\fbinv.
Only the single-lepton final state is considered, where the lepton is either an electron or a muon.
No significant excess of data events over the standard model expectation is observed. Upper limits on the $\PX_{5/3}$ pair production cross section are set at 95\% confidence level (\CL) using a multivariate discriminant. 

The $\PX_{5/3}$ masses up to 1630\GeV have been excluded at 95\% \CL in the decay mode to a top quark and a \PW boson.
These are the most restrictive limits to date in this $\PX_{5/3}$ decay mode.
In the decay mode to a top quark and a charged Higgs boson, $\PX_{5/3}$ masses up to 1740\GeV are excluded at 95\% \CL, with the strongest exclusion obtained for an \PHpm mass of 800\GeV. 
For the $\PHpm$ mass range studied, $\PX_{5/3}$ masses between 1620 and 1740\GeV are excluded at 95\% CL.
The interpretation in the $\PX_{5/3} \to \PQt\PSHp$ decay chain is restricted to simulated mass points satisfying the kinematic condition that the $\PX_{5/3}$ mass exceeds the sum of the charged Higgs boson and top quark masses, and no interpolation of signal efficiencies or templates between mass hypotheses is performed.
These are the first limits on the $\PX_{5/3}$ decaying into a top quark and a charged Higgs boson.

\begin{acknowledgments}
We congratulate our colleagues in the CERN accelerator departments for the excellent performance of the LHC and thank the technical and administrative staffs at CERN and at other CMS institutes for their contributions to the success of the CMS effort. In addition, we gratefully acknowledge the computing centers and personnel of the Worldwide LHC Computing Grid and other centers for delivering so effectively the computing infrastructure essential to our analyses. Finally, we acknowledge the enduring support for the construction and operation of the LHC, the CMS detector, and the supporting computing infrastructure provided by the following funding agencies: SC (Armenia), BMFWF and FWF (Austria); FNRS and FWO (Belgium); CNPq, CAPES, FAPERJ, FAPERGS, and FAPESP (Brazil); MES and BNSF (Bulgaria); CERN; CAS, MoST, and NSFC (China); MINCIENCIAS (Colombia); MSES and CSF (Croatia); RIF (Cyprus); SENESCYT (Ecuador); ERC PRG and PSG, TARISTU24-TK10 and MoER TK202 (Estonia); Academy of Finland, MEC, and HIP (Finland); CEA and CNRS/IN2P3 (France); SRNSF (Georgia); BMFTR, DFG, and HGF (Germany); GSRI (Greece); MATE and NKFIH (Hungary); DAE and DST (India); IPM (Iran); SFI (Ireland); INFN (Italy); MSIT and NRF (Republic of Korea); MES (Latvia); LMTLT (Lithuania); MOE and UM (Malaysia); BUAP, CINVESTAV, CONACYT, LNS, SEP, and UASLP-FAI (Mexico); MOS (Montenegro); MBIE (New Zealand); PAEC (Pakistan); MSHE, NSC, and NAWA (Poland); FCT (Portugal);  MESTD (Serbia); MICIU/AEI and PCTI (Spain); MOSTR (Sri Lanka); Swiss Funding Agencies (Switzerland); NSTC (Taiwan); MHESI (Thailand); TUBITAK and TENMAK (T\"{u}rkiye); NASU (Ukraine); STFC (United Kingdom); DOE and NSF (USA).

\begin{sloppypar}
\setlength\emergencystretch{\hsize}
\hyphenation{Rachada-pisek} Individuals have received support from the Marie-Curie program and the European Research Council and Horizon 2020 Grant, contract Nos.\ 675440, 724704, 752730, 758316, 765710, 824093, 101115353, 101002207, 101001205, and COST Action CA16108 (European Union); the Leventis Foundation; the Alfred P.\ Sloan Foundation; the Alexander von Humboldt Foundation; the Science Committee, project no. 22rl-037 (Armenia); the Fonds pour la Formation \`a la Recherche dans l'Industrie et dans l'Agriculture (FRIA) and Fonds voor Wetenschappelijk Onderzoek contract No. 1228724N (Belgium); the Beijing Municipal Science \& Technology Commission, No. Z191100007219010, the Fundamental Research Funds for the Central Universities, the Ministry of Science and Technology of China under Grant No. 2023YFA1605804, the Natural Science Foundation of China under Grant No. 12535004, and USTC Research Funds of the Double First-Class Initiative No.\ YD2030002017 (China); the Ministry of Education, Youth and Sports (MEYS) of the Czech Republic; the Shota Rustaveli National Science Foundation (Georgia); the Deutsche Forschungsgemeinschaft (DFG), among others, under Germany's Excellence Strategy -- EXC 2121 ``Quantum Universe" -- 390833306, and under project number 400140256 - GRK2497; the Hellenic Foundation for Research and Innovation (HFRI), Project Number 2288 (Greece); the Hungarian Academy of Sciences, the New National Excellence Program - \'UNKP, the NKFIH research grants K 131991, K 138136, K 143460, K 143477, K 147557, K 146913, K 146914, K 147048, TKP2021-NKTA-64, 2026-4.1.1-MISSZI\'{O}K-2026-00005 and 2025-1.1.5-NEMZ\_KI-2025-00004, and MATE KKP and KKPCs Research Excellence and Flagship Research Groups grants (Hungary); the Council of Science and Industrial Research, India; ICSC -- National Research Center for High Performance Computing, Big Data and Quantum Computing, FAIR -- Future Artificial Intelligence Research, and CUP I53D23001070006 (Mission 4 Component 1), funded by the NextGenerationEU program, the Italian Ministry of University and Research (MUR) under Bando PRIN 2022 -- CUP I53C24002390006, PRIN PRIMULA 2022RBYK7T (Italy); the Latvian Council of Science; the Ministry of Science and Higher Education, project no. 2022/WK/14, and the National Science Center, contracts Opus 2021/41/B/ST2/01369, 2021/43/B/ST2/01552, 2023/49/B/ST2/03273, and the NAWA contract BPN/PPO/2021/1/00011 (Poland); the Funda\c{c}\~ao para a Ci\^encia e a Tecnologia (Portugal); the National Priorities Research Program by Qatar National Research Fund; MICIU/AEI/10.13039/501100011033, ERDF/EU, ``European Union NextGenerationEU/PRTR", projects PID2022-142604OB-C21, PID2022-139519OB-C21, PID2023-147706NB-I00, PID2023-148896NB-I00, PID2023-146983NB-I00, PID2023-147115NB-I00, PID2023-148418NB-C41, PID2023-148418NB-C42, PID2023-148418NB-C43, PID2023-148418NB-C44, PID2024-158190NB-C22, RYC2021-033305-I, RYC2024-048719-I, CNS2023-144781, CNS2024-154769 and Plan de Ciencia, Tecnolog{\'i}a e Innovaci{\'o}n de Asturias, Spain; the Chulalongkorn Academic into Its 2nd Century Project Advancement Project, the National Science, Research and Innovation Fund program IND\_FF\_68\_369\_2300\_097, and the Program Management Unit for Human Resources \& Institutional Development, Research and Innovation, grant B39G680009 (Thailand); the Eric \& Wendy Schmidt Fund for Strategic Innovation through the CERN Next Generation Triggers project under grant agreement number SIF-2023-004; the Kavli Foundation; the Nvidia Corporation; the SuperMicro Corporation; the Welch Foundation, contract C-1845; and the Weston Havens Foundation (USA).
\end{sloppypar}
\end{acknowledgments}\section*{Data availability} Release and preservation of data used by the CMS Collaboration as the basis for publications is guided by the  \href{https://doi.org/10.7483/OPENDATA.CMS.1BNU.8V1W}{CMS data preservation, re-use and open access policy}.

\bibliography{auto_generated} 
\cleardoublepage \appendix\section{The CMS Collaboration \label{app:collab}}\begin{sloppypar}\hyphenpenalty=5000\widowpenalty=500\clubpenalty=5000\cmsinstitute{Yerevan Physics Institute, Yerevan, Armenia}
{\tolerance=6000
A.~Gevorgyan\cmsorcid{0000-0003-2751-9489}, A.~Hayrapetyan, V.~Makarenko\cmsorcid{0000-0002-8406-8605}, A.~Tumasyan\cmsAuthorMark{1}\cmsorcid{0009-0000-0684-6742}
\par}
\cmsinstitute{Institut f\"{u}r Hochenergiephysik, Vienna, Austria}
{\tolerance=6000
P.S.~Hussain\cmsorcid{0000-0002-4825-5278}, M.~Sonawane\cmsorcid{0000-0003-0510-7010}
\par}
\cmsinstitute{Marietta Blau Institute for Particle Physics, Vienna, Austria}
{\tolerance=6000
W.~Adam\cmsorcid{0000-0001-9099-4341}, L.~Benato\cmsorcid{0000-0001-5135-7489}, T.~Bergauer\cmsorcid{0000-0002-5786-0293}, M.~Dragicevic\cmsorcid{0000-0003-1967-6783}, S.~Gundacker\cmsorcid{0000-0003-2087-3266}, A.K.~Guven\cmsorcid{0009-0004-5670-5138}, M.~Jeitler\cmsAuthorMark{2}\cmsorcid{0000-0002-5141-9560}, N.~Krammer\cmsorcid{0000-0002-0548-0985}, A.~Li\cmsorcid{0000-0002-4547-116X}, D.~Liko\cmsorcid{0000-0002-3380-473X}, M.~Matthewman, A.~Pfeiffer\cmsorcid{0000-0001-5328-448X}, J.~Schieck\cmsAuthorMark{2}\cmsorcid{0000-0002-1058-8093}, R.~Sch\"{o}fbeck\cmsAuthorMark{2}\cmsorcid{0000-0002-2332-8784}, M.~Shooshtari\cmsorcid{0009-0004-8882-4887}, N.~Van~Den~Bossche\cmsorcid{0000-0003-2973-4991}, W.~Waltenberger\cmsorcid{0000-0002-6215-7228}, C.E.~Wulz\cmsAuthorMark{2}\cmsorcid{0000-0001-9226-5812}
% ...
\par}
$^{1}$Also at Yerevan State University, Yerevan, Armenia\\
$^{2}$Also at Technische Universit\"{a}t Wien, Vienna, Austria\\
% ...
\end{sloppypar}
%%% END EDITABLE REGION %%%
% skeleton_end
\end{document}